\documentclass[aps,pre,twocolumn,superscriptaddress,showpacs]{revtex4-1}
\usepackage{graphicx}
\usepackage{dcolumn}
\usepackage{bm}
\usepackage{epsfig}
\usepackage{color}

\begin{document}


\title{A Tetris-like Model Showing a Universal Enhanced Flow Rate of a
  Hopper Discharging Hard Discs Through an Adjustable Inclusion}


\author{Guo-Jie Jason Gao}
\email{koh.kokketsu@shizuoka.ac.jp, gjjgao@gmail.com}
\affiliation{Department of Mathematical and Systems Engineering,
  Shizuoka University, Hamamatsu, Shizuoka 432-8561, Japan}

\date{\today}

\begin{abstract}
In the literature, placing an inclusion near the orifice of a hopper,
containing disc particles, has been experimentally and numerically
reported to locally enhance the gravity-driven hopper flow
rate. Moreover, the peaked flow rate can happen regardless of the
interparticle friction, the inclusion geometry, or the disc
dispersity. To reveal the fundamental reason causing this local
effect, we propose a Tetris-like model that sequentially moves one
disc particle at a time towards the hopper orifice. A Gaussian
displacement function that independently controls a disc's movement in
the horizontal or vertical direction, and the algorithm of the model
accepts a movement as long as it creates no overlap between objects in
the system. Our model creates an artificial steady probability-driven
hopper flow without knowing the Newtonian dynamics which allows
interparticle collaborative motion. Under specific conditions, we
reproduce the enhanced flow rate and show that a moderate response
time of the system and a flow rate difference between its value around
the inclusion and its maximum without an inclusion are sufficient to
explain this local effect with no Newton's laws involved.
\end{abstract}


\maketitle

\section{Introduction}
\label{introduction}

Multiple phases including gas, liquid and solid can coexist within
short length scale comparable to particle size in nonequilibrium
athermal systems such as a symmetric slot hopper continuously
discharging granular particles under gravity. The empirical Beverloo
equation captures the behavior of the funnel flow where only a part of
particles near the center of the hopper flows, $Q \sim \rho \sqrt g
W_o^{{5 \mathord{\left/ {\vphantom {5 2}} \right.
      \kern-\nulldelimiterspace} 2}}$, where $Q$ is the mass flow
rate, $\rho$ is the bulk density of the granular medium, $g$ is
gravity and $W_o$ is the orifice width of the hopper. On the other
hand, the Johanson equation, $Q \sim \rho \sqrt g W_o^{{3
    \mathord{\left/ {\vphantom {3 2}} \right.
      \kern-\nulldelimiterspace} 2}}$, describes the mass flow where
almost all particles inside the hopper move simultaneously
\cite{beverloo61, johanson65, janda12, jaeger92, hutter94, jaeger96,
  campbell06, jop06, thomas13}.

Moreover, to globally increase the gravity-driven hopper flow rate,
introducing more isotropic particles \cite{baxter89, cleary02, wu06,
  cleary08, fraige08, hohner12, thomas13}, increasing the particle
dispersity \cite{potapov96, denniston99, remy11}, or placing an
inclusion near the hopper orifice \cite{tuzun85, yang01} has been
shown effective. The last strategy has also been shown experimentally
and numerically to locally enhance the hopper flow rate of frictional
particles recently, that is, the flow rate exhibits a local peak as an
inclusion that creates a pressure reduction is placed at an optimal
height above the hopper orifice \cite{zuriguel11, lozano12,
  alonso-marroquin12}. However, this local effect cannot be explained
by conventional continuum laws based on the assumption of constant
density for hopper flows mentioned above.

\begin{figure}
\includegraphics[width=0.45\textwidth]{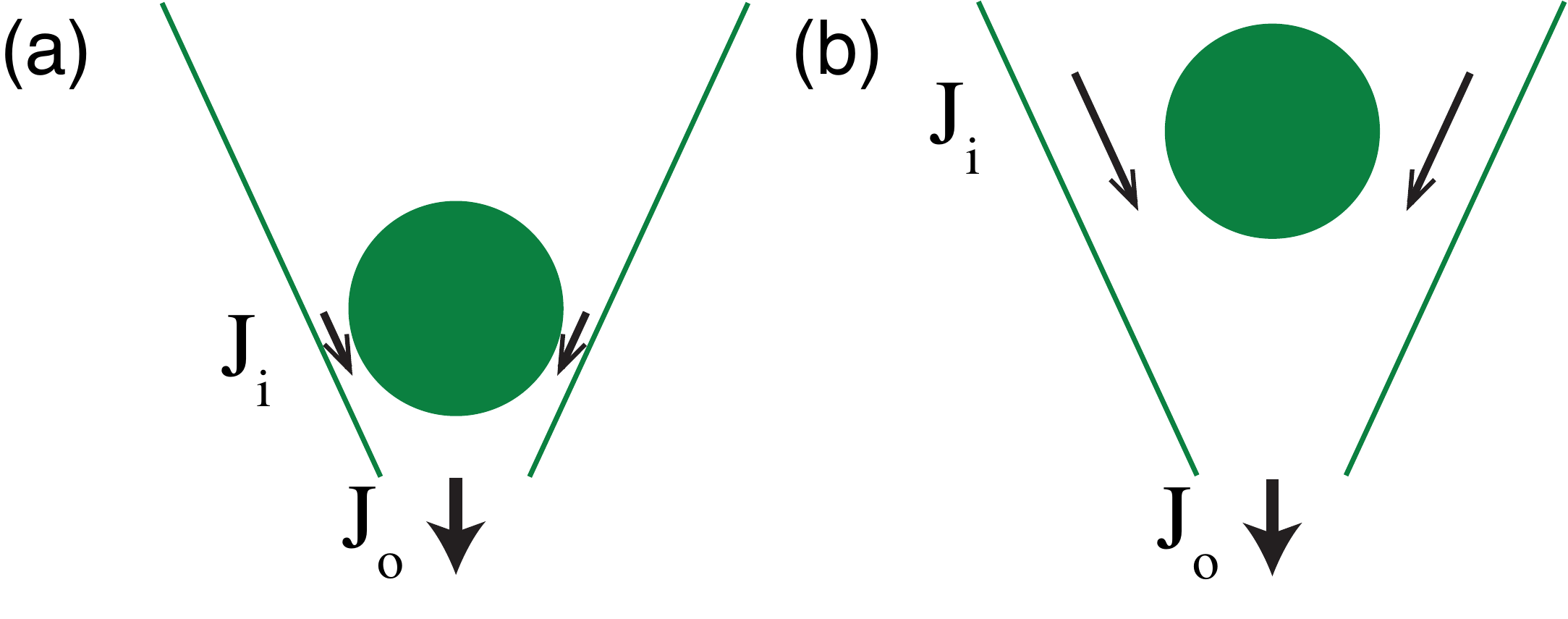}
\caption{\label{fig:hypothesis} (Color online) Schematic of a hopper
  (solid lines) containing an inclusion (circle). $J_i$ is the hopper
  flow rate at the inclusion height, with the part of the hopper lower
  than the center of the inclusion sliced off, and $J_o$ is the
  maximal flow rate while the hopper contains no inclusion. The hopper
  flow can be either (a) fluidized with $J_i<J_o$ or (b) clogging with
  $J_i>J_o$.}
\end{figure}

In our recent study of frictionless particles interacting via the
pairwise linear spring potential using molecular dynamics (MD)
simulation, we also find the flow rate can be locally peaked and show
that this local feature survives regardless of the interparticle
friction, the particle size dispersity and the inclusion geometry
\cite{gao18}. The interparticle interactions do determine where the
local flow rate happens. For example, a frictional system tends to
exhibit its local flow rate peak with the inclusion placed higher than
a frictionless system. In the light of our MD results, we have
narrowed down possible reasons causing the locally enhanced flow rate
to two candidates: 1) the interparticle collaborative motion due to
the Newtonian dynamics and 2) the flow rate difference between its
value $J_i$ at the inclusion height above the hopper orifice and its
maximum $J_o$ while the hopper contains no inclusion, as shown
schematically in Fig. \ref{fig:hypothesis}, where (a) and (b) depict a
fluidized flow regime with $J_i<J_o$ and a clogging flow regime with
$J_i>J_o$, respectively. The local peak of the actual flow rate $J_a$
occurs because it can be boosted during the hopper flow transits from
fluidized to clogging while $J_i$ becomes greater than $J_o$. To find
out whether the Newtonian dynamics contributes indispensably to this
local phenomenon, we propose a purely geometrical model that closely
resembles the classical video game \textit{Tetris}, where objects of
various shapes fall down the playing field one at a time and can move
left or right freely with a prescribed step size during the falling
process without creating any overlap, except all objects are uniform
circles in this study. A similar but more simplified Tetris-like model
that contains rectangular objects occupying a 45-degree tilted square
lattice and allows additional upward moving of objects had been used
to study granular compaction under vibration \cite{nicodemi97}. Our
Tetris-like model produces an probability-driven hopper flow that can
clog due to transient arching and particles blocking one another
during position update with no Newtonian dynamics involved. Our
results show that the local flow rate peak still exists under specific
conditions, which serves as a decisive evidence that the Newtonian
dynamics is not essential for this local phenomenon to happen.

Below we elaborate on our Tetris-like model, generating an artificial
probability-driven hopper flow in section \ref{method}, followed by
quantitative analysis of the flow rates with or without an inclusion
in section \ref{results_and_discussions} and a mathematical model
describing this local effect. We conclude our study in section
\ref{conclusions}.

\section{Numerical simulation method}
\label{method}

\begin{figure}
\includegraphics[width=0.40\textwidth]{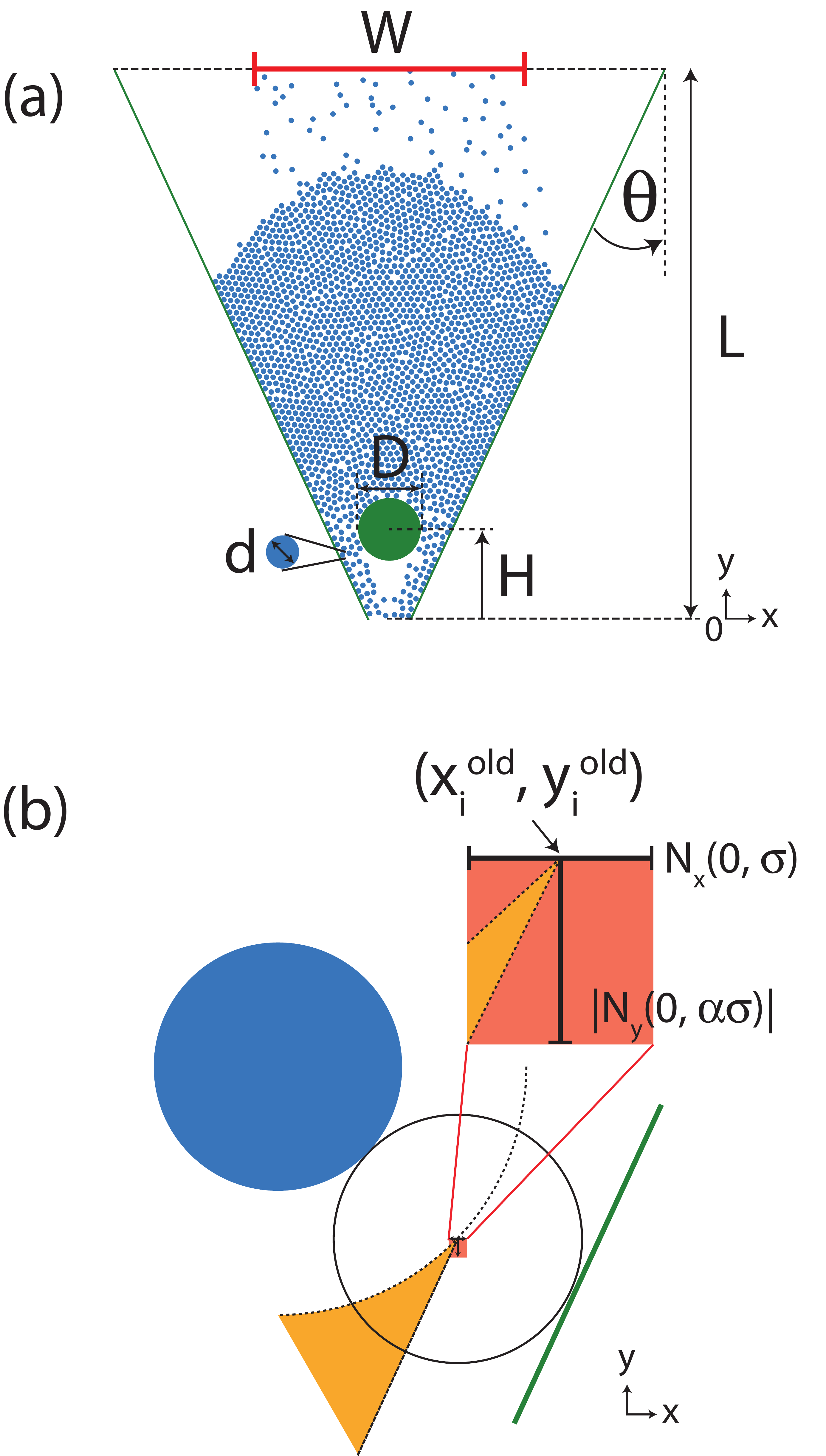}
\caption{\label{fig:gaussian_MC_model} (Color online) (a) The
  simulation setup of a symmetric hopper (slant solid lines) of equal
  height and top-width $L$ and hopper angle $\theta$. The hopper
  discharges disc particles (small circles) of diameter $d$ and
  possessing horizontal (x) and vertical (y) positions, which reenter
  the hopper from its top border, centered ($x=0$) within $W =
  0.5L$. An inclusion (large circle) of diameter $D$ sits at a height
  $H$ above the hopper orifice ($y=0$). (b) The probabilistic
  Tetris-like model, exemplarily demonstrated by a particle (open
  circle) in between another particle (solid circle) and a wall (slant
  solid line). The fanned area (shaded) depicts a region where the
  sandwiched particle can move into at the next position update,
  governed by two independent Gaussian functions, $N_x$ and $N_y$ with
  zero means and standard deviations $\sigma$ and $\alpha \sigma$,
  respectively. $\sigma = 0.05d$ and $\alpha$ is a variable in this
  study.}
\end{figure}

We develop a Tetris-like model to study the probability-driven hopper
flow of monodisperse disc particles of diameter $d$, discharged from a
geometrically symmetric hopper with a height $L=83d$ and a hopper
angle $\theta=0.4325$ radians, as shown schematically in
Fig. \ref{fig:gaussian_MC_model}(a). A circular inclusion of diameter
$D=0.112L$ is placed on the symmetric axis of the hopper and away from
its orifice by a controllable distance $H$. The size ratio between the
inclusion and a disc particle is $D/d=9.296$.

Within each position update in our Tetris-like model, each particle
$i$ has exact one chance to change its horizontal (x) and vertical (y)
positions from $(x_i^{old}, y_i^{old})$ to $(x_i^{new}, y_i^{new})$,
based on
\begin{equation} \label{eqn_x}
x_i^{new} = x_i^{old} + {N_x}(0,\sigma),
\end{equation}
and
\begin{equation} \label{eqn_y}
y_i^{new} = y_i^{old} + \left| {{N_y}(0,\alpha \sigma)} \right|,
\end{equation}
where the two independent Gaussian functions, $N_x$ has a zero mean
and a standard deviation $\sigma$, while $N_y$ also has a zero mean
but a standard deviation $\alpha\sigma$. In both functions,
$\sigma=0.05d$, as shown exemplarily in
Fig. \ref{fig:gaussian_MC_model}(b). A larger value of $\alpha$, a
variable in this study, represents a stronger trend of particles
moving downwards to the orifice of the hopper, similar to discharging
animate or inanimate particles through constrictions with increasing
driving force \cite{zuriguel15}. We take the absolute value of $N_y$
so that particles move only towards the hopper orifice with no
backward movement. The algorithm of the Tetris-like model accepts a
change of position of particle $i$ if it creates no overlap between
the particle and any other objects in the system. Otherwise, the
attempted position change will be discarded and particle $i$ stays
unmoved. We update the positions of particles sequentially based on a
random order that is different for each position update.

To measure the hopper flow rate $J_a$ while the inclusion is placed at
a given height $H$ above the hopper orifice, we initiate a simulation
with randomly arranged particles. There are $N=2048$ initially
randomly placed particles in the system, which can fill the hopper up
to about $2/3$ of its height when the system reaches a steady state of
a probability-driven hopper flow. To maintain a constant number of
particles $N$, a particle dropping out of the hopper will reenter the
system from the hopper's top border with its vertical (y) position
artificially shifted by a distance $L$ and its horizontal (x) position
reassigned randomly within $W \in [ - L/4,L/4]$. The latter strategy
is to maintain a steady hopper flow without particles piling up to the
top border of the hopper, an undesirable boundary effect that affects
the hopper flow rate. We then wait for $10,000$ position updates so
that the system becomes fully relaxed from the initial condition and
reaches a steady probability-driven hopper flow. After that, we count
the number of particles passing the hopper orifice within $990,000$
position updates. For each value of $H$, we use 50 different initial
conditions to calculate the average and the variance of the actual
flow rate $J_a$ leaving the orifice in terms of number of particles
leaving the hopper per update. We define $J_o$ as the value of $J_a$
while the hopper contains no inclusion. We also measure $J_i$, as the
number of particles flowing through the internal passages between the
inclusion and the hopper walls on its both sides per update, at the
same height $H$ as the center of the inclusion is located above the
orifice of the hopper in an identical way, except we put particles
dropping below height $H$ back to the top of the hopper. This approach
is essentially the same as measuring $J_i$ by slicing off the part of
the hopper below the center of the inclusion so that the removed piece
of hopper has no effect on $J_i$.

\begin{figure}
\includegraphics[width=0.40\textwidth]{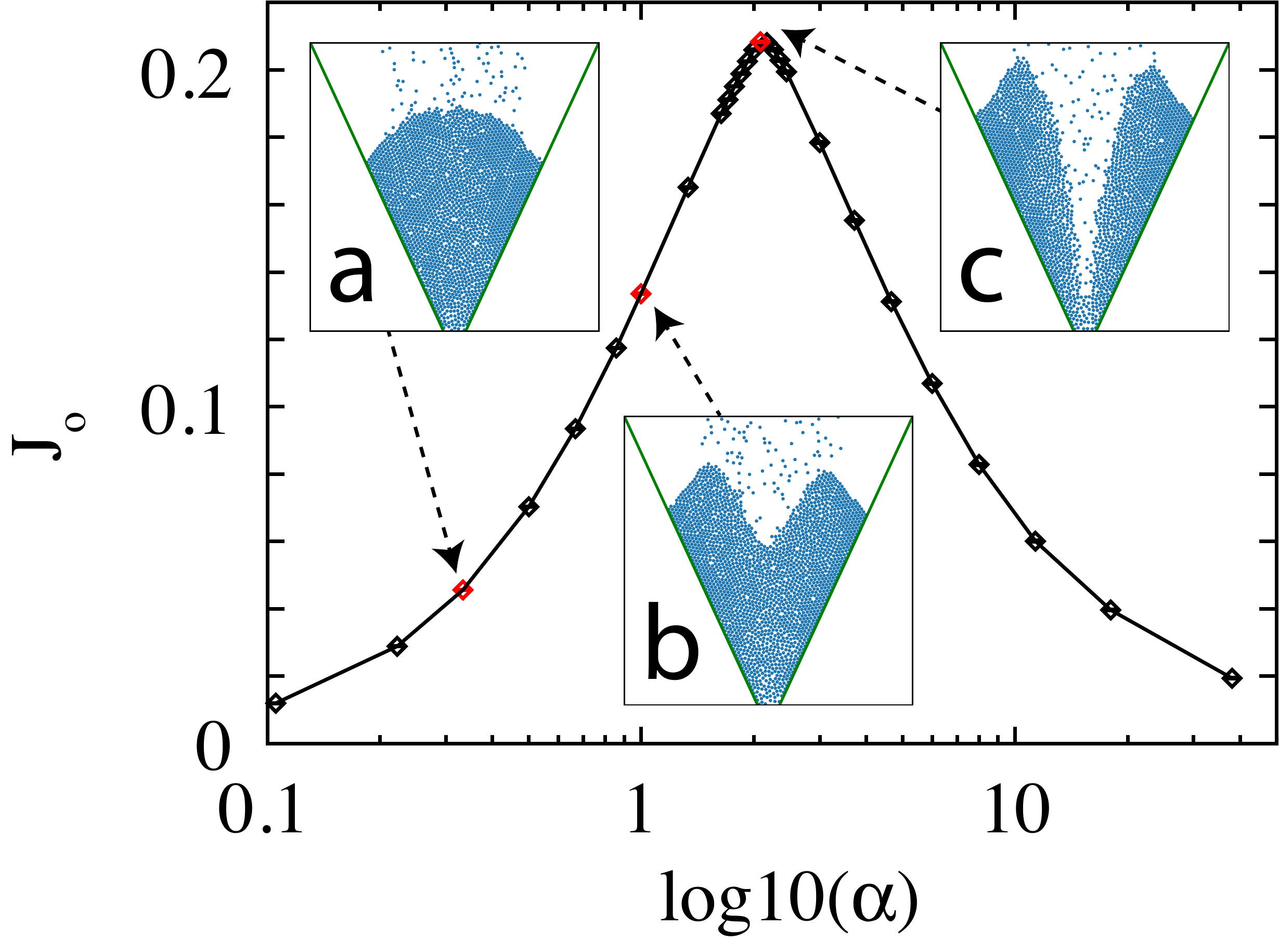}
\caption{\label{fig:gaussian_orifice_flowrates} (Color online) Hopper
  flow rate $J_o$ as a function of $log_{10}(\alpha)$ while the hopper
  contains no inclusion. The error bars, smaller than the symbols in
  the plot, are obtained using $50$ initial conditions. The insets
  show exemplary snapshots of $\alpha$ = (a) 1/3, (b) 1.0, and (c)
  2.082 (diamonds), respectively.}
\end{figure}

\begin{figure}
\includegraphics[width=0.39\textwidth]{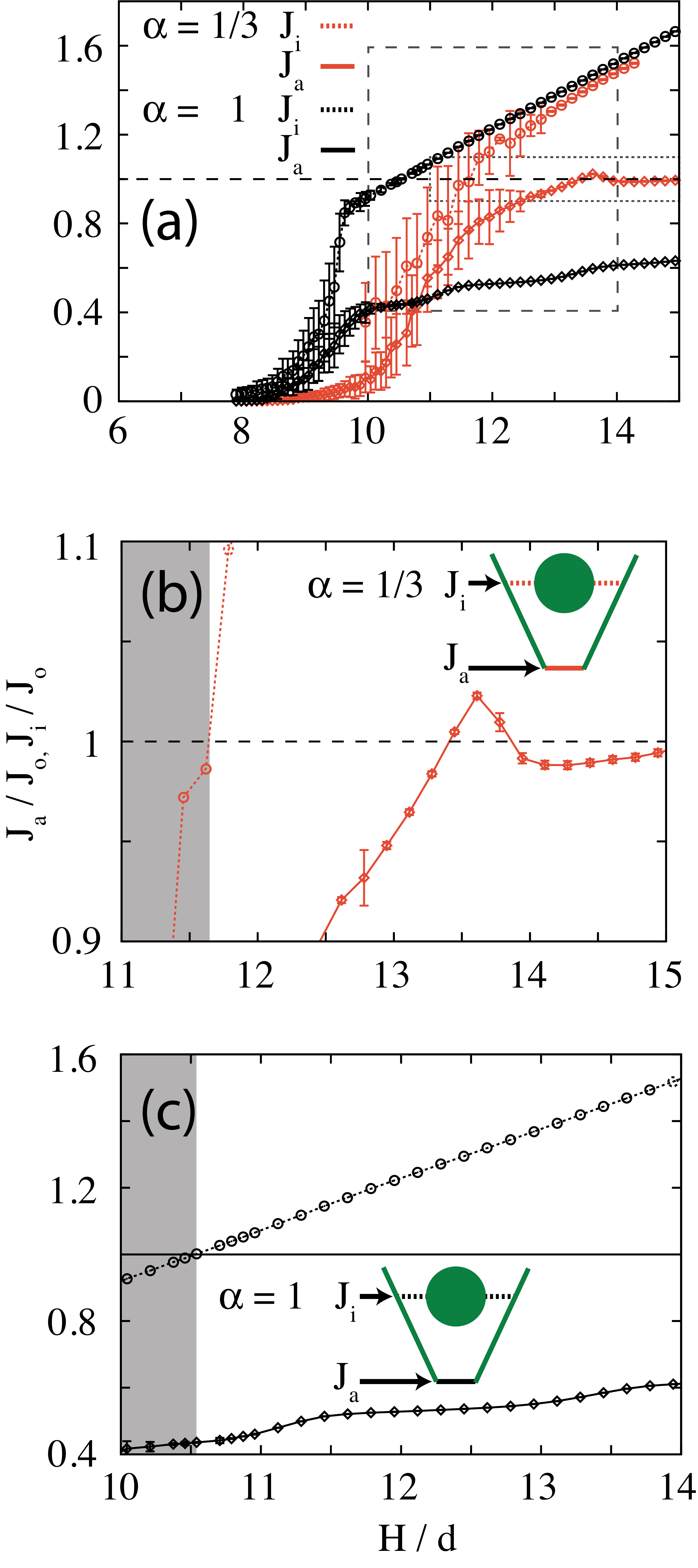}
\caption{\label{fig:gaussian_normalized_flowrates_all} (Color online)
  (a) Hopper flow rates $J_i$ measured at the inclusion height (dotted
  lines with circles) and $J_a$ measured at the hopper orifice (solid
  lines with diamonds) with $\alpha=1/3$ (the 2nd and the 3rd curves
  from top) and $1.0$ (the 1st and the 4th curves from top),
  respectively. The error bars, most of them smaller than the symbols
  in the plot, are obtained using $50$ initial conditions. (b) and (c)
  show the zoomed plots of $J_i$ and $J_a$ in the dotted region of
  $\alpha=1/3$ and the dashed region of $\alpha=1.0$ in (a),
  separately. The fluidized flow regimes, where $J_i<J_o$ shown in
  Fig. \ref{fig:hypothesis}(a), are shaded.}
\end{figure}

\section{Results and Discussions}
\label{results_and_discussions}
To understand the behavior of the probability-driven hopper flow while
the hopper contains no inclusion, we measured $J_o$ as a function of
$\alpha$. The results are shown in
Fig. \ref{fig:gaussian_orifice_flowrates}. $J_o$ initially increases
monotonically with $\alpha$ until reaches its maximum at $\alpha
\approx 2.082$. As can been seen in the insets of
Fig. \ref{fig:gaussian_orifice_flowrates}, in the monotonically
increasing regime of $J_o$, the hopper flow performs a free surface
like a bloated hump, a crater and a deeply-carved pit as $\alpha$
changes from $1/3$, $1.0$ to $2.082$, respectively. $J_o$ then
decreases to zero as $\alpha$ approaches infinity, where the rejection
rate of relocating a particle becomes too high and a steady hopper
flow is impossible. To avoid potential stagnant effect, we avoid this
negatively-sloped regime of large $\alpha$ while studying the behavior
of $J_i$ and $J_a$ when the hopper contains an inclusion.

Next, to verify our hypothesis that a flow rate difference between
$J_i$ and $J_o$ may be sufficient to create a locally enhanced flow
rate of $J_a$ in our Tetris-like model that creates a
probability-driven hopper flow without the Newtonian dynamics, we
measured $J_i$ and $J_a$, normalized by $J_o$, as a function of
$H/d$. The results are shown in
Fig. \ref{fig:gaussian_normalized_flowrates_all}(a). We chose
$\alpha=1/3$ and $1.0$, which give a low and a medium flow rates,
respectively, in the positively sloped regime of $J_o(\alpha)$, as
discussed in Fig. \ref{fig:gaussian_orifice_flowrates} above. We can
see clearly in the zoomed
Fig. \ref{fig:gaussian_normalized_flowrates_all}(b) of $\alpha=1/3$
that soon after $J_i$ becomes higher than $J_o$ at $H/d \approx 11.6$,
$J_a$ exhibits a local peak at $H/d \approx 13.6$. The local peak can
be explained by a flow rate difference between $J_i$ at the inclusion
height $H$ and $J_o$, together with an overall slower response time of
the system, due to a smaller $\alpha$ and therefore a lower flow
rate. The amount of particles passing the internal passages between
the inclusion and the hopper walls while $J_i$ is moderately higher
than $J_o$ is able to reach the hopper orifice before the system can
response to it and limit $J_a$ to be no greater than $J_o$. As a
result, $J_a > J_o$ is possible. As we lift the inclusion further away
from the orifice, a flow of rate higher than $J_o$ has to travel
longer to reach the hopper orifice and therefore has a higher chance
to clog. The net result of this competition is a disappearing enhanced
effect and eventually the value of $J_a$ saturates at $J_o$ as the
inclusion sits too high to affect the flow rate $J_a$. On the other
hand, although there still exists a greater than unity $J_i/J_o$ after
$H/d > 10.5$ in the zoomed
Fig. \ref{fig:gaussian_normalized_flowrates_all}(c) of a larger
$\alpha=1.0$, we do not observe the enhanced effect. In this case, as
$J_i$ become higher than $J_o$, the system always responses fast
enough to effectively limit the actual flow rate $J_a$ out of the
orifice below $J_o$.

\begin{figure}
\includegraphics[width=0.37\textwidth]{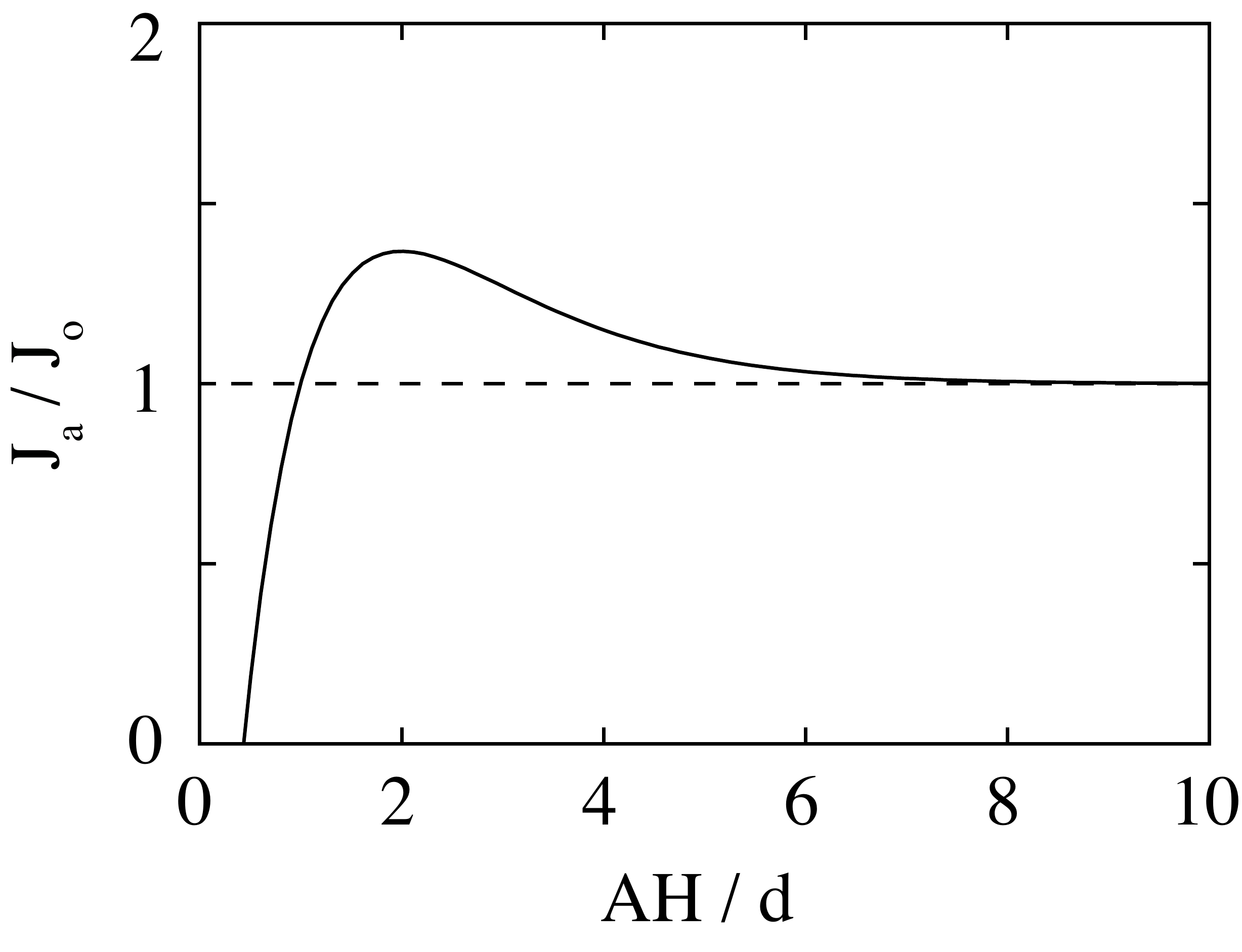}
\caption{\label{fig:gaussian_flowrates_math_model} Predicted $J_a/J_o$
  of hopper flow, where $AH_o/d=1.0$ and $B/A=1.0$, described by the
  qualitative mathematical formula of Eqn. (\ref{math_model}) with the
  same scaling factors $A$ and $B$.}
\end{figure}

Finally, we propose a simplified mathematical formula that
qualitatively captures the flow mechanism described above.  With a
fixed value of $\alpha$, the flow rate $J_a$ leaving the hopper should
be proportional to the strength of ${J_i} \sim {{A(H - {H_0})}
  \mathord{\left/ {\vphantom {{A(H - {H_0})} d}} \right.
    \kern-\nulldelimiterspace} d}$ at the inclusion height, where $A$
is a scaling factor, and $H_0$ is the height of the inclusion when
particles in the hopper start to flow. $J_i$ grows linearly as a
function of $H/d$ based on our simulation results, as shown in
Fig. \ref{fig:gaussian_normalized_flowrates_all}(b) and (c). As $J_i$
increases with the inclusion placed at a higher $H$, more and more
particles can pass the two passages between the inclusion and the
hopper walls, and the probability $P_c$ of particle clogging below the
inclusion grows as well. We assume ${P_c} = {e^{\left( {{B
        \mathord{\left/ {\vphantom {B A}} \right.
          \kern-\nulldelimiterspace} A}} \right){J_i}}}$, where $B$ is
another scaling factor, and $J_a$ is inversely proportional to
$P_c$. Together, we obtain ${J_a} \sim {{{J_i}} \mathord{\left/
    {\vphantom {{{J_i}} {{P_c}}}} \right.  \kern-\nulldelimiterspace}
  {{P_c}}}$, which can be explicitly expressed as
\begin{equation} \label{math_model}
{J_a} = {J_o}\left[ {1 + \frac{{{{A(H - {H_0})} \mathord{\left/
 {\vphantom {{A(H - {H_0})} d}} \right.
 \kern-\nulldelimiterspace} d}}}{{{e^{\left( {{B \mathord{\left/
 {\vphantom {B A}} \right.
 \kern-\nulldelimiterspace} A}} \right)\left( {{{A(H - {H_0})} \mathord{\left/
 {\vphantom {{A(H - {H_0})} d}} \right.
 \kern-\nulldelimiterspace} d}} \right)}}}}} \right].
\end{equation}
A term of $1$ is included so that $J_a$ approaches $J_o$ as the
inclusion is placed at an $H$ far away from the hopper orifice and has
no effect on $J_a$. Fig. \ref{fig:gaussian_flowrates_math_model} shows
an exemplary hopper flow rate described by this formula. The actual
forms of the linear relation, ${J_i} \sim {{A(H - {H_0})}
  \mathord{\left/ {\vphantom {{A(H - {H_0})} d}} \right.
    \kern-\nulldelimiterspace} d}$, and the assumed exponential term,
${P_c} = {e^{\left( {{B \mathord{\left/ {\vphantom {B A}} \right.
          \kern-\nulldelimiterspace} A}} \right){J_i}}}$, in
Eqn. (\ref{math_model}), however, are to be investigated in our next
study.

\section{Conclusions}
\label{conclusions}

In principle, the interparticle friction, the cooperative motion
between particles and the geometry of grains and inclusion play
important roles in granular hopper flow. For example, they determine
precisely where the local peak of the hopper flow rate occurs, as
previously reported in a gravity-driven frictional system, where disc
particles passing through a round inclusion. Our previous study of
frictionless particles using MD simulations has shown that altering
the interparticle friction, the particle dispersity or the inclusion
geometry affects the position of the local flow rate peak but does not
necessarily eliminate it \cite{gao18}. To investigate the importance
of the long-range interparticle cooperative motion due to the
Newtonian dynamics, we use a purely geometrical Tetris-like model that
produces a probability-driven hopper flow and completely switch off
the Newtonian interaction. Surprisingly, we can successfully reproduce
the locally enhanced flow rate, and our results clearly show that a
flow rate difference between its value $J_i$ at the inclusion height
$H$ and its maximum $J_o$ while the hopper contains no inclusion,
together with a slow response time of the system due to a moderate
downward moving trend controlled by $\alpha$ in Eqn. (\ref{eqn_y}), is
sufficient to qualitatively reproduce this local phenomenon. As a
result, the Newtonian dynamics is not indispensable for observing this
local effect. This study serves as an example of deciphering a
perplexing phenomenon in an athermal granular system by reducing its
dynamics to the minimal to reveal the fundamental factor playing under
the surface.

\section*{acknowledgments}
GJG gratefully acknowledges financial support from Shizuoka University
startup funding. GJG also thanks Corey S. O'Hern for useful
discussions.

\bibliography{68890}

\end{document}